# Present and Future Galaxy Redshift Surveys:

## ORS, DOGS and 2dF

Ofer Lahav[1]

*Institute of Astronomy, Madingley Rd., Cambridge CB3 0HA, UK*

**Abstract.** Three galaxy redshifts surveys and their analyses are discussed. (i) The recently completed Optical Redshift Survey (ORS) includes galaxies larger than 1.9 arcmin and/or brighter than $14.5^m$. It provides redshifts for $\sim 8300$ galaxies at Galactic latitude $|b| > 20°$. A new analysis of the survey explores the existence and extent of the Supergalactic Plane (SGP). Its orientation is found to be in good agreement with the standard SGP coordinates, and suggests that the SGP is at least as large as the survey (16000 km/sec in diameter). (ii) The Dwingeloo Obscured Galaxy Survey is aimed at finding galaxies hidden behind the Milky-Way using a blind search in 21 cm. The discovery of Dwingeloo1 illustrates that the survey will allow us to systematically survey the region $30° < l < 200°$ out to 4000 km/sec. (iii) The Anglo-Australian 2-degree-Field (2dF) survey will yield 250,000 redshifts for APM-selected galaxies brighter than $19.5^m$ to map the large scale structure on scales larger than $\sim 30\,h^{-1}$ Mpc. To study morphological segregation and biasing the spectra will be classified using Artificial Neural Networks.

## 1. Introduction

Redshift surveys in the last decade provided a major tool for cosmographical and cosmological studies. In particular, surveys such as CfA, SSRS, IRAS and Las Campanas yielded useful information on local structure and on cosmological parameters such as $\Omega$ from redshift distortion and from comparison with the peculiar velocity field (for review see e.g. Dekel 1994 ; Strauss & Willick 1995). Together with measurements of the Microwave Background Radiation and gravitational lensing the redshift surveys provide a major probe of the world geometry and the dark matter. In particular, to understand the issues of morphological segregation and biasing it is important to select the samples at different wavelengths. In this review we focus on 3 surveys: the optically selected ORS and 2dF and the DOGS at 21cm.

---

[1]email: lahav@ast.cam.ac.uk

## 2. The Optical Redshift Survey (ORS)

The Optical Redshift Survey (ORS, Santiago et al. 1995) is a redshift survey covering the sky at Galactic latitude $|b| > 20°$. The survey was drawn from the UGC, ESO and ESGC catalogues, and it contains two subsets: one complete to a blue magnitude of 14.5, and the other complete to a blue major diameter of 1.9 arcmin. The entire sample consists of 8457 galaxies. Redshifts are now available to 98 % of them, and about 1300 new redshifts were measured to complete the survey. The selection function of the survey is rather complicated due to the diversity of catalogues used and the Galactic extinction (Santiago et al. 1996). In fact, a weight is associated with each galaxy to compensate for the incompleteness at a given direction and distance. The high number density of galaxies ($n \sim 0.1$ galaxies per $(h^{-1}\,\mathrm{Mpc})^3$) in ORS makes it ideal for cosmographical studies of the local universe. A preliminary analysis of the correlation function in redshift space yields $\xi(s) \approx (s/7.5\,h^{-1}\,\mathrm{Mpc})^{-1.8}$ for the magnitude limited ORS (Hermit et al., in preparation).

For the purpose of studying whole-sky aspects (e.g. the SGP below) we filled in the Zone of Avoidance (ZOA) at $|b| < 20°$ with galaxies from the IRAS 1.2Jy survey (Fisher et al. 1995a). The Zone of Avoidance in IRAS $|b| < 5°$ was filled in by interpolation based on the observed galaxy distribution below and above the ZOA (Yahil et al. 1991). Hereafter when we refer to the ORS sample we mean the ORS sample ($|b| > 20°$) supplemented by the IRAS 1.2Jy galaxies at $|b| < 20°$. The weight of each IRAS galaxies (the inverse of the selection function) is multiplied by the ratio of ORS/IRAS number density, a factor of $\sim 2.5$. For comparison we also apply our analysis to the complete IRAS 1.2Jy (Fisher et al. 1995a) which includes 5313 galaxies over 87.6 % of the sky.

Figure 1 shows the distributions of ORS and IRAS galaxies (constructed as described above) in a sphere of radius of 4000 km/sec, projected on the standard Supergalactic coordinates. The SGP is in visible 'edge-on' in the $(SGZ - SGX)$ projection. We note that the structures seen in ORS and IRAS are quite similar, but the ORS map is much denser, and clusters are more prominent.

## 3. The Supergalactic Plane Revisited

The ORS and IRAS surveys allow us to revisit the Supergalacctic Plane. The following is based on recent work by Lahav et al. (in preparation)

### 3.1. A Brief History of the SGP

The so-called Supergalactic Plane (SGP) was recognized by de Vaucouleurs (1956) using the Shapley-Ames catalogue, following an earlier analysis of radial velocities of nearby galaxies which suggested a differential rotation of the 'metagalaxy' by Vera Rubin. This remarkable feature in the distribution of nebulae was in fact already noticed by William Herschel more than 200 years ago.

Although this feature is clearly visible in whole-sky galaxy catalogues the SGP has only been little re-examined quantitatively in recent years. Tully (1986) claimed the flattened distribution of clusters extends across a diameter of $\sim 0.1c$ with axial ratios of 4:2:1. Shaver & Pierre (1989) found that radio galaxies are more strongly concentrated to the SGP than are optical galaxies, and that the SGP as represented by radio galaxies extends out to redshift $z \sim 0.02$. Di Nella & Paturel (1995) recently revisited the SGP using a compilation of nearly 5700 galaxies larger than 1.6 arcmin, and found qualitatively agreement with the standard SGP.

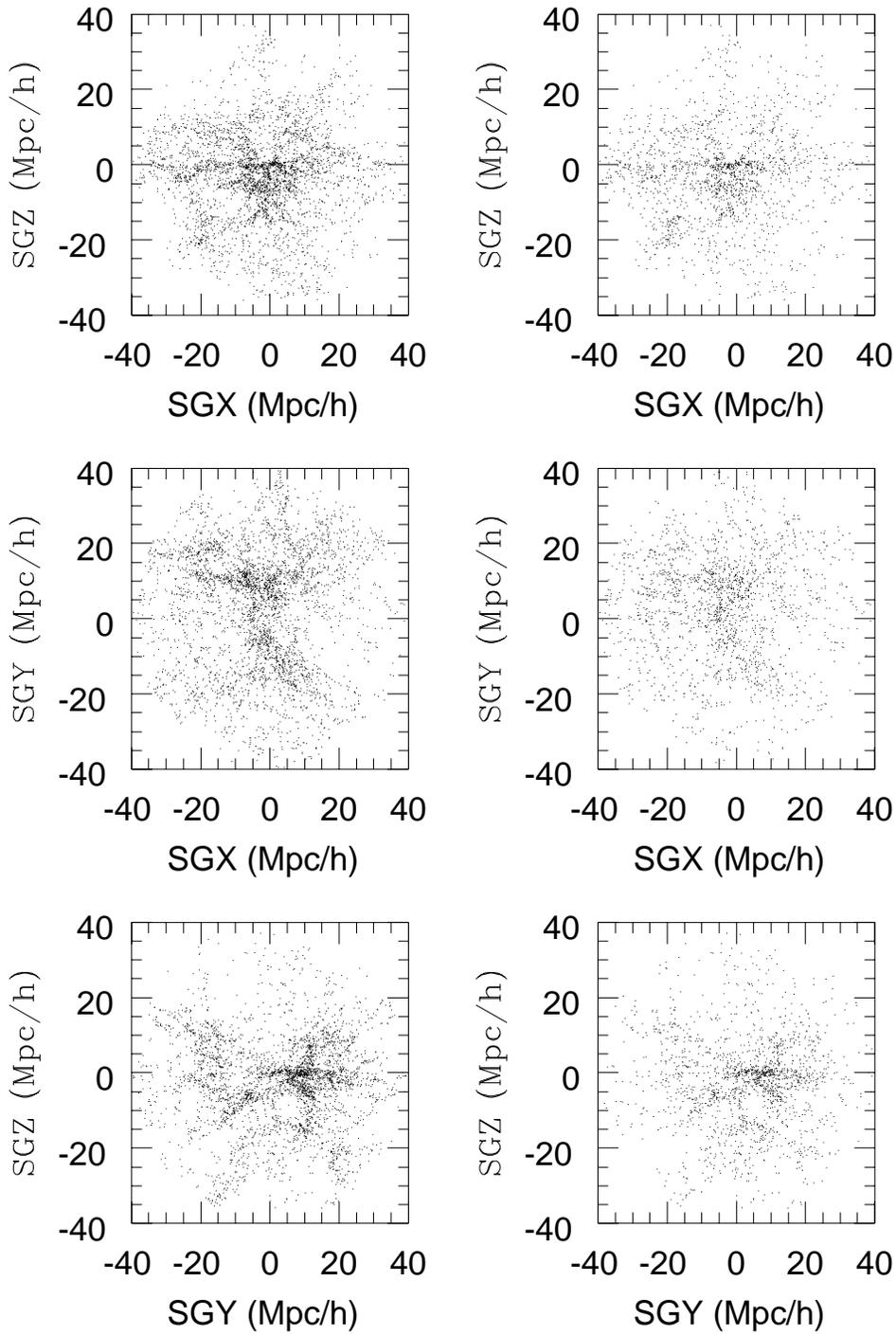

Figure 1. The distribution of galaxies in ORS (left) and in IRAS (right) projected within a sphere of radius of 4000 km/sec in standard Supergalactic coordinates.

The currently used coordinates for the SGP (de Vaucouleurs et al. 1976) in which the equator lies along the SGP have the north pole at the direction of Galactic coordinates ($l = 47.37°; b = +6.32°$). The origin of SGL is at ($l = 137.37°, b = 0°$), which is one of the two regions where the SGP is crossed by the Galactic Plane. The Virgo cluster is at SGP coordinates ($SGL = 104°; SGB = -2°$). Traditionally the Virgo cluster was regarded as the centre of the Supergalaxy, and this was termed the 'Local Supercluster'. But recent maps of the local universe indicate that much larger clusters, such as the Great Attractor and Perseus-Pisces on opposite sides of the Local Group are major components of this 'plane'. Supergalactic coordinates are commonly used in extragalactic studies, but the degree of linearity, extent and direction of the SGP have been little quantified in recent years. Moreover, it is important to compare the extent of the SGP with other filamentary structures seen in redshift surveys and in $N$-body simulations. Top-down scenarios (e.g. Hot Dark Matter) in particular predict the formation of Zeldovich pancakes, although these are also seen in hierarchical (bottom-up) scenarios (e.g. Cold Dark Matter).

### 3.2. Moment of Inertia Analysis

Let's imagine that we wish to detect a slab embedded in a uniform sphere of radius $R$. One approach is to construct a covariance matrix ('moment of inertia tensor') for the *fluctuation* in the density field:

$$\tilde{C}_{ij} = C_{ij} - \bar{C}_{ij} = \frac{1}{N_{eff}} \sum_{gal} \frac{1}{\phi(|\mathbf{x}|)} (x_i - \bar{x}_i)(x_j - \bar{x}_j) - \delta_{ij} \frac{n_{bg}}{<n>} \frac{R^2}{5} . \qquad (1)$$

where $N_{eff}$ is the effective number of galaxies corrected for the radial selection function $\phi$, and $x_i, x_j$ ($i, j = 1, 2, 3$) are Cartesian components of $\mathbf{x}$, $n_{bg}$ is the *background* density in the absence of the slab, and $<n>$ the mean density (*including* the slab). The last term is due to a uniform distribution with density $n_{bg}$. The next step is to diagonalize the covariance matrix and to find the eigen-values and eigen-vectors :

$$\tilde{C} \mathbf{u}_\alpha = \lambda_\alpha \mathbf{u}_\alpha, \qquad (2)$$

where the $\lambda_\alpha$'s and $\mathbf{u}_\alpha$'s are the eigen-values and eigen-vectors ($\alpha = 1, 2, 3$). The 'half-width' ('1-sigma') along each of the 3 axes is given by $\sqrt{\lambda_\alpha}$. Note that since the background contribution (the last term in eq. 1) is isotropic, it only affects the eigen-values, but not the directions of the eigen-vectors. This procedure is essentially a Principal Component Analysis (PCA), a well known statistical tool for reducing dimensionality of parameter space, here from 3 to 2 dimensions.

### 3.3. The orientation and size of the SGP

To use eq. (1) we need to estimate the centre $\bar{\mathbf{x}}$ and mean densities $<n>$ and $n_{bg}$. It is interesting to note that the 'centre' moves away from the origin (us) by no more than 800 km/sec for ORS and 600 km/sec for IRAS. This is partially due to the 'tug of war' between the Great Attractor and Perseus-Pisces. Indeed the 'centre' moves back towards the origin for $R_{max} > 4000$ km/sec. As a crude estimate we calculated the mean number density 'outside' the SGP for $|SGB| > 20^o$. We adopt as fiducial values $n_{bg} = 0.085$ and $0.035$ galaxies per $(h^{-1}\,\text{Mpc})^3$ for ORS and IRAS respectively. The global number densities, including the SGP region, are $<n> \sim 0.1$ and $\sim 0.04$ galaxies per $(h^{-1}\,\text{Mpc})^3$ for ORS and IRAS, respectively. Having estimated the 'centre' and the mean background density we can now construct the moment of inertia tensor (eq. 1) and diagonalize it (eq. 2). As we found the 'centre' to be close to us at the origin we choose hereafter $\bar{x}_i = 0$.

The PCA indeed finds one axis to be much smaller than the other two, i.e. it identifies objectively a 'plane'. To quantify the relation between this plane and the standard SGP coordinates we focus on the angle $\theta_z$ between the normal to the standard SGP and the normal to our PCA-identified plane. There are variations of $\theta_z$ with $R_{max}$ but typically $\theta_z \sim 30^o$, also when calculated for differential shells. The probability of 2 vectors to be within an angle $\theta_z$ by chance is $P(<\theta_z) = 1 - \cos(\theta_z)$, i.e. $\sim 13\%$. The joint probability for several shells to be aligned by chance to the observed value is given by the product of the probabilities for each of the shells, giving here only few percents.

Given the close alignment with the standard SGP coordinates, we shall assume these axes with us at the origin, and calculate the variances $\lambda_\alpha$'s along SGX, SGY and SGZ, as a function of $R_{max}$, after subtracting the mean background term (eq. 1). We then define $a = \sqrt{\lambda_1}$, $b = \sqrt{\lambda_2}$ and $c = \sqrt{\lambda_3}$ as the semi-axes (1-sigma) of the 'ellipsoid'. One problem in deriving $a, b$ and $c$ is the uncertainty in the mean background density $n_{bg}$ which results in some cases in negative $\lambda_\alpha$'s. In these cases we set $\lambda_\alpha = 0$. The variation of $a, b$ and $c$ with $R_{max}$ shows several features. The main one is that while the semi-minor axis $c$ (corresponding to the $SGZ$ axis) levels off (in IRAS) or even declines (in ORS) with $R_{max}$, the larger axes $a$ and $b$ (corresponding to $SGX$ and $SGY$) grow out to the sample boundary $R_{max} = 8000$ km/sec. This indicates that we can only place a *lower* limit of the size of the SGP : we do not actually see the end of the SGP ! The analysis also suggests that optical galaxies are more concentrated towards the SGP than IRAS galaxies. e.g. at $R_{max} = 6000$ km/sec the axial ratios are $a : b : c = 4.8 : 3.7 : 1.0$ for ORS and $2.9 : 1.6 : 1.0$ for IRAS.

It is worth recalling that PCA is only a suitable analysis if there is a single plane-like feature. Clearly, in the real universe there is 'contamination' due to clusters and voids below and above the SGP which may 'confuse' the PCA. We examined the smoothed density field and found the region $|SGB| < 20^o$ to be far denser than the rest of the volume. It is remarkable that the 'blind' PCA finds the SGP in spite of this background contamination problem, indicating that indeed the SGP is the major feature in the local universe.

### 3.4. Wiener reconstruction of the IRAS SGP

Two major problems affect most studies of redshift surveys: shot noise and redshift distortion. One approach to deal with both is Wiener filtering (Fisher et al 1995b). In particular let us expand the density field $\delta(\mathbf{r}) = \frac{\delta\rho}{\rho}$ in spherical

harmonics:
$$\rho(\mathbf{r}) = \sum_l \sum_m \sum_n C_{ln}\, \rho_{lmn}\, j_l(k_n r)\, Y_{lm}(\hat{\mathbf{r}}), \qquad (3)$$

where the $k_n$'s are chosen e.g. to satisfy the boundary condition that the logarithmic derivative of the potential is continuous at $r = R_{max}$. An estimator of the coefficients from the redshift data is

$$\hat{\rho}^S_{lmn} = \sum_{gal} \frac{1}{\phi(s)}\, j_l(k_n s)\, Y^*_{lm}(\hat{\mathbf{r}}). \qquad (4)$$

Fisher et al. (1995b) showed that the real-space coefficients can be reconstructed by

$$\hat{\delta}^R_{lmn} = \sum_{n'n''} \left(\mathbf{S}_l\, [\mathbf{S}_l + \mathbf{N}_l]^{-1}\right)_{nn'} \left(\mathbf{Z}_l^{-1}\right)_{n'n''} \hat{\delta}^S_{lmn''}, \qquad (5)$$

where the matrix $\mathbf{Z}_l$ (which depends on the assumed $\Omega^{0.6}/b$) corrects for redshift distortion, and $\mathbf{S}_l\, [\mathbf{S}_l + \mathbf{N}_l]^{-1}$ is the Wiener matrix which filters the data where they are noisy. The signal matrix $\mathbf{S}_l$ depends on the assumed prior power-spectrum of fluctuations. It can be shown that this gives the optimal reconstruction in the minimum variance sense, and it can also be derived from Bayesian arguments and Gaussian probability distribution functions. In this approach the density field goes to the mean density at large distances. This does not mean necessarily that the SGP itself disappears at large distances, it only reflects our ignorance on what exists far away, where the data are very poor.

A nice property of this Wiener approach applied to the density field $\frac{\delta\rho}{\rho}$ is that it will also give the optimal reconstruction for any property which is linear in $\frac{\delta\rho}{\rho}$. In particular, if we seek the optimal reconstruction of the moment of inertia

$$\tilde{C}_{ij} = C_{ij} - \bar{C}_{ij} = \frac{1}{N} \int\int\int [\rho(\mathbf{x}) - n_{bg}](x_i - \bar{x}_i)(x_j - \bar{x}_j)\, dV, \qquad (6)$$

it can be expressed *analytically* (by substituting eq. 3 into eq. 6) in terms of the reconstructed coefficients $\delta^R_{lmn}$. The full mathematical details will be given elsewhere (Webster, Fisher and Lahav, in preparation). Here we shall simply report on an application of the technique to the IRAS 1.2Jy out to $R_{max} = 20000$ km/sec (assuming as priors $\Omega^{0.6}/b = 1$, $\sigma_8 = 0.7$ and a CDM power spectrum with $\Omega h = 0.2$, and using 6905 $\delta_{lmn}$ coefficients). The PCA applied to the Wiener reconstructed moment of inertia tensor gives a misalignment angle $\theta_z = 16°$ (compared with $\sim 30°$ in the analysis above). It is interesting that the agreement with the standard SGP is getting even better when corrections are made for the redshift distortion and the shot-noise.

## 4. The Dwingeloo Obscured Galaxy Survey (DOGS)

The disk of the Milky Way contains gas and dust which obscures about 20% of the optical extragalactic sky, the so-called "Zone of Avoidance" (ZOA). Nearby galaxies hidden behind the ZOA may have an important influence on the dynamics of the Local Group and its peculiar motion relative to the cosmic microwave background radiation. It is also important to probe the ZOA in order to verify the connectivity of large scale features such as the SGP discussed above. However, such galaxies suffer extinction by dust and gas at optical wavelengths, and confusion by stars in the infrared. Emission at 21cm by neutral atomic hydrogen (HI) associated with late-type galaxies may be observed if the velocity of the emission differs from that of the local gas (Kerr & Henning 1987). Thus, it is possible to detect in HI galaxies behind the ZOA which are very difficult to detect at other wavelengths.

### 4.1. The DOGS strategy

Utilising the 25m Dwingeloo radio-telescope, the DOGS team (Burton, Ferguson, Henning, Kraan-Korteweg, Lahav, Loan & Lynden-Bell) is systematically surveying almost the whole of the Northern Galactic Plane ($30° < l < 200°$) below a Galactic latitude $|b| \leq 5°$. The nearly 15000 pointings are ordered in a honeycomb pattern with a spacing of $0.4°$. The receiver bandwidth of 20 MHz gives coverage over the velocity range $0 < V_{LSR} < 4000$ km/sec (in the frame of the Local Standard of Rest, LSR). Lower velocities have already been surveyed by the Dwingeloo 25m as part of the Dwingeloo/Leiden Galactic HI survey. The strategy was to first conduct a fast search with 5 minute integrations to uncover the most massive or nearby galaxies, followed by a full 1 hour integration per pointing to produce a deeper, flux-limited catalogue. The fast blind search has been completed and the deeper 1hr search continues.

### 4.2. Dwingeloo1 and neighbours

In the preliminary reduction of some of the 5 min 'fast' blind search data, Kraan-Korteweg and Loan noticed in August 1994 an interesting 'double horned' signal in several neighbouring pointings of the telescope. We named this new galaxy Dwingeloo1 (hereafter Dw1). For more details and HI spectrum of this galaxy see Kraan-Korteweg et al. (1994). The position of Dw1 (at Galactic coordinates $l \approx 138.5°; b \approx -0.1°$) coincided with a feature 2.2 arcmin in diameter previously noted by Hau et al. (1994) on a red Palomar Sky Survey plate.

Observations with the Westerbork Synthesis Radio Telescope (WSRT) confirmed the reality of Dw1. Optical $V, R, I$ imaging by the Isaac Newton Telescope and by the Wise Observatory clearly revealed Dw1, with a distinct bar and spiral arms, suggesting an SBb morphology. The foreground extinction in the blue is estimated to be about 5.8 magnitudes, so much of the outermost parts of Dw1 may remain hidden.

The recession velocity of Dw1 is $V_{LSR} = +110$ km/sec. Correcting for the motion within our Milky Way, the radial velocity relative to our Galaxy is $V_G \approx 256$ km/sec. The (corrected) rotation velocity of Dw1 is 113 km/sec. Using the INT I,V, R magnitudes and the Tully-Fisher relation (calibrated on the global Hubble flow) it was recently estimated (Loan et al. 1996) that the distance to Dw1 is about 300 km/sec. The major source of uncertainty in the distance ($\pm 200$ km/sec) is due to the poorly known extinction at such low Galactic latitude.

Most likely, Dw1 belongs to the IC342 group (which includes IC342, Maffei I, Maffei II and other smaller galaxies), at the outskirts of the Local Group. We intend to further explore this group, and to study its effect on the motions of the Local Group and nearby galaxies. Three close neighbours of Dw1 were found since its discovery. A detailed Westerbork map (Burton et al. 1996) revealed a new galaxy, Dwingeloo2, 20 arcmin away from Dw1. Dwingeloo2's HI diameter is about half that of Dw1. Recently two more possible dwarf companions of Maffei I were discovered by McCall & Buta (1995). The discovery of these 4 galaxies probably does not change our basic picture of the dynamics, mass and motion of the Local Group. However, it illustrates that more surprises are expected behind the Milky Way. Future radio surveys, e.g. the 13-beam receiver on the 64m radiotelescope in Parkes Australia, will continue to probe the ZOA. Other useful probes are in the infrared (e.g. IRAS and the DENIS and 2MASS surveys in 2 $\mu$) and in X-ray (e.g. ROSAT).

## 5. The 2 degree Field (2dF) Survey

Existing optical and IRAS redshift surveys contain 10,000-20,000 galaxies. A major step forward using multifibre technology will allow in the near future to produce redshift surveys of millions of galaxies. In particular, there are two major surveys on the horizon. The American Sloan Digital Sky Survey (SDSS) will yield images in 5 colours and redshifts for about 1 million galaxies over a quarter of the sky in the northern Galactic cap. (Gunn and Weinberg 1995). It will be carried out using a dedicated 2.5m telescope in New Mexico. The median redshift of the survey is $z \sim 0.1$.

A complementary Anglo-Australian survey, the 2 degree Field (2dF) will produce redshifts for 250,000 galaxies brighter than $b_J = 19.5^m$ (with median redshift of $z \sim 0.1$), selected from the APM survey. The survey will utilize a new 400-fibre system on the 4m AAT, covering $\sim 1,700$ sq deg of the sky. About 250,000 spectra will be measured over $\sim 100$ nights. This large collaboration (Cole, Collins, Dalton, Efstathiou, Ellis, Folkes, Frenk, Lahav, Maddox, Peacock, Sutherland from the UK; and Bland-Hawthorne, Cannon, Colless, Couch, Driver, Glazebrook, Kaiser, Lumsden, Peterson, Taylor from Australia) will produce $\sim 12,500$ redshifts per collaborator ! The survey will probe scales larger than $\sim 30\, h^{-1}$ Mpc, allowing to fill in the gap between scales probed by previous local galaxy surveys and the scale of $\sim 1000\, h^{-1}$ Mpc probed by COBE. It will also allow accurate determination of $\Omega$ and bias parameter from redshift distortion. A deeper extension down to $R = 21$ for 10,000 galaxies is also planned.

### 5.1. Automated Spectral Classification of Galaxies

Surveys like 2df and SDSS will produce unusually large numbers of galaxy spectra, providing an important probe of the intrinsic galaxy properties. The integrated spectrum of a faint galaxy is an important measure of its stellar composition as well as its dynamical properties. Moreover, spectral properties correlate fairly closely with morphology. Indeed, as the spectra are more directly related to the underlying astrophysics, they are a more robust classifier for evolutionary and environmental probes. Spectra can be obtained to larger redshifts than ground-based morphologies and, as 1-D datasets, are easier to analyse. Although the concept of spectral classification goes back to Humason, Morgan & Mayall, few uniform data sets are available and most contain only a small number of galaxies (e.g. Kennicutt 1992). Recent spectral analyses for classification were out carried by Francis et al. (1992) for QSO spectra, von-Hippel et al. (1994) for stellar spectra, and in particular for galaxy spectra by Sodré & Cuevas (1994), Heyl (1994), and Connolly et al.(1994),

Spectral classification is important for several practical and fundamental reasons. In order to derive luminosities corrected for the effects of redshift, the $k$-correction must be estimated for each galaxy. The rest-frame spectral energy distribution is needed, which can be obtained by matching the observed spectrum against templates of local galaxies. The proportion of sources in each class as a function of luminosity and redshift is of major interest. Apart from its relevance for environmental and evolutionary studies, new classes of objects may be discovered as outliers in spectral parameter space. Furthermore, by incorporating spectral features with other parameters (e.g. colour and velocity dispersion) an 'H-R diagram for galaxies' can be examined with possible important implications for theories of galaxy formation.

The challenge is to design a computer algorithm which will reproduce classification to the same degree a student or a colleague of the human expert can do it. Such an automated procedure usually involves two steps: (i) feature extraction from a digitised image or a galaxy spectrum e.g. by Principal Component Analysis (PCA). (ii) A classification procedure, in which a computer 'learns' from a 'training set' for which a human expert provided his or her classification.

PCA allows to reduce the dimensionality of the input parameter space. By identifying the linear combination of input parameters with maximum variance, PCA finds the Principal Components that can be most effectively used to characterize the inputs. Apart from being a data compression technique, PCA is in fact an example of 'unsupervised learning', in which an algorithm or a linear 'network' discovers features and patterns for itself.

Artificial Neural Networks (ANNs), originally suggested as simplified models of the human brain, are computer algorithms which provide a convenient general-purpose framework for classification (Hertz et al. 1991). ANNs are related to other statistical methods common in Astronomy and other fields. In particular ANNs generalise Bayesian methods, multi-parameter fitting, PCA, Wiener filtering and regularisation methods (e.g. Lahav 1994 for a summary). Consider now the specific problem of morphological classification of galaxies. If the type is $T$ (e.g. on de Vaucouleurs' numerical system [-6,11]) and we have a set of parameters $\mathbf{x}$ (e.g. diameters, colours or spectral features) then we would like to find free parameters $\mathbf{w}$ ('weights') such that

$$\sigma^2 = \frac{1}{N_{gal}} \sum_i [T_i - f(\mathbf{w}, \mathbf{x_i})]^2, \quad (7)$$

(where the sum is over the galaxies), is minimized. The function $f(\mathbf{w}, \mathbf{x})$ is the 'network'. Commonly $f$ is written in terms of non-linear functions. For a given Network architecture the first step is the 'training' of the ANN. In this step the weights are determined by minimizing 'least-squares' (e.g. eq. 7) e.g. by the Backpropagation or Quasi-Newton algorithms.

Recently the method was applied to a sample of 830 APM galaxy images (Naim et al. 1995; Lahav et al. 1995) by extracting features directly from the images, and training the net on the human classification from 6 experts. A dispersion of 1.8 $T$-units, was achieved when the ANN was trained and tested on the mean type as deduced from all available expert classifications. There was a remarkable similarity in the dispersion between two human experts and that between ANN and experts. In other words, the results indicated that the ANNs can replicate the expert's classification of the APM sample as well as other colleagues or students of the expert can do.

Using this combination of PCA and ANN for galaxy spectra (Folkes & Lahav 1995), it was shown in a pilot study that galaxy spectra from Kennicutt's sample can still be classified when degraded to the 2dF set-up. The spectra can typically be compressed using only 8 Principal Components. By training ANN on the spectra with known morphological type $T$ (eq. 7) it was shown that spectra with signal/noise ratio corresponding to the 2dF limiting magnitude of $19.5^m$ can still be correctly classified into one of 5 classes with success rate of over 85 %.

## 6. Discussion

The three surveys described here illustrate the rapid growth in systematic redshift surveys. Increasing the number of redshifts by a factor of 100 does not necessarily mean that our cosmological inference will be 100 times larger. However, the new deep survey will dramatically improve our statistics for studying important issues such as the word geometry and the nature of the dark matter. They also call for the development of new tools, such as ANN, in exploring galaxy parameters in large data sets.

**Acknowledgments.** I thank all my collaborators for their contribution to the work presented here, in particular B. Santiago & M. Strauss (ORS), A. Loan & R. Kraan-Korteweg (DOGS), and S. Folkes (2dF).